\documentclass[11pt]{article}
\usepackage{amssymb, amsmath, amsthm, color, cite} 
\usepackage[breaklinks,colorlinks=false]{hyperref} 

\newcommand{\thmref}[1]{\hyperref[#1]{{Theorem~\ref*{#1}}}}
\newcommand{\lemref}[1]{\hyperref[#1]{{Lemma~\ref*{#1}}}}
\newcommand{\corref}[1]{\hyperref[#1]{{Corollary~\ref*{#1}}}}
\newcommand{\eqnref}[1]{\hyperref[#1]{{(\ref*{#1})}}}
\newcommand{\defref}[1]{\hyperref[#1]{{Definition~\ref*{#1}}}}
\newcommand{\secref}[1]{\hyperref[#1]{{Section~\ref*{#1}}}}
\newcommand{\propref}[1]{\hyperref[#1]{{Proposition~\ref*{#1}}}}
\newcommand{\claimref}[1]{\hyperref[#1]{{Claim~\ref*{#1}}}}

\newtheorem{theorem}{Theorem}
\newtheorem{lemma}{Lemma}
\newtheorem{corollary}{Corollary}
\newtheorem{definition}{Definition}

\newtheorem{claim}{Claim}
\newtheorem{remark}{Remark}

\newcommand{\tensor}{\otimes}
\newcommand{\abs}[1]{\lvert #1\rvert}

\newcommand{\ket}[1]{\lvert #1 \rangle}
\newcommand{\bra}[1]{\langle #1 \vert}
\def\ketbra #1#2{\ket{#1}\!\bra{#2}}

\newcommand{\tr}{{\mathsf{Tr}}}

\newcommand{\Good}{{\mathsf{Good}}}

\newcommand{\M}{{\mathcal M}}
\newcommand{\A}{{\mathcal A}}

\newcommand{\poly}{\mathrm{poly}}
\newcommand{\negl}{\mathrm{negl}}

\newcommand{\cX}{{\mathcal X}}
\newcommand{\cY}{{\mathcal Y}}
\newcommand{\cZ}{{\mathcal Z}}

\newcommand{\cS}{{\mathcal S}}

\newcommand{\bH}{{\mathbb H}}
\newcommand{\BPP}{\mathsf{BPP}}
\newcommand{\BQP}{\mathsf{BQP}}
\newcommand{\ZK}{\mathsf{ZK}}
\newcommand{\QSZK}{\mathsf{QSZK}}
\newcommand{\QCZK}{\mathsf{QCZK}}
\newcommand{\QIP}{\mathsf{QIP}}

\newcommand{\IP}{\mathsf{IP}}
\newcommand{\AM}{\mathsf{AM}}
\newcommand{\QAM}{\mathsf{QAM}}
\newcommand{\V}{{\mathcal V}}
\renewcommand{\P}{{\mathcal P}}

\newcommand{\comment}[1]{\emph{\color{red}Comment:\color{black} #1}} 
\newlength{\commentslength}
\newcommand{\comments}[1]{
\hspace{-2\parindent}
\addtolength{\commentslength}{-\commentslength}
\addtolength{\commentslength}{\linewidth}
\addtolength{\commentslength}{-\parindent}
\fcolorbox{red}{white}{\smallskip\begin{minipage}[c]{\commentslength}
\emph{Comments:}\begin{itemize}#1\end{itemize}\end{minipage}}\bigskip
}
\renewcommand{\comment}[1]{}\renewcommand{\comments}[1]{}

\def \AMbr {\langle \A, \M \rangle}
\def\na {{n_1}}
\def\nb {{n_2}}
\def\nc {{n_c}}
\def\a {A} 
\def\b {B} 
\def\c {C} 
\def\h {H} 
\def\f {F} 
\def\argmax{{\arg\,\max}}
\def\Ex {\mathrm{E}}

\begin{document}

\author{
Rahul Jain\thanks{Centre for Quantum Technologies and Department of Computer Science, National University of Singapore,%
~\tt{rahul@comp.nus.edu.sg}}
\and Alexandra Kolla\thanks{EECS Department, Computer Science Division, University of California,%
~\tt{\{akolla, gatis\}@cs.berkeley.edu}}
\and Gatis Midrij\=anis$^\dagger$
\and Ben W. Reichardt\thanks{School of Computer Science Department and Institute for Quantum Computing, University of Waterloo,%
~\tt{breic@iqc.ca}}
}

\title{On parallel composition of zero-knowledge proofs with black-box quantum simulators}

\date{}
\maketitle
\thispagestyle{empty}
\abstract{
Let $L$ be a language decided by a constant-round quantum Arthur-Merlin ($\QAM$) protocol with negligible soundness error and all but possibly the last message being classical.  We prove that if this protocol is zero knowledge with a black-box, quantum simulator $\cS$, then $L \in \BQP$.  Our result also applies to any language having a three-round quantum interactive proof ($\QIP$), with all but possibly the last message being classical,  with negligible soundness error and a black-box quantum simulator.   

These results in particular make it unlikely that certain protocols can be composed in parallel in order to reduce soundness error, while maintaining zero knowledge with a black-box quantum simulator.  They generalize analogous classical results of Goldreich and Krawczyk (1990).  

Our proof goes via a reduction to quantum black-box search. We show that the existence of a black-box quantum simulator for such protocols when $L \notin \BQP$ would imply an impossibly-good quantum search algorithm.
}
\newpage
\setcounter{page}{1}

\section{Introduction}

A {\em zero-knowledge} (ZK) protocol for language $L$ allows a prover
to convince a verifier the membership of an input $x$ in $L$, without
disclosing any extra information.  That is when $x \in L$, anything
{\em efficiently} computable after interacting with the prover could
also have been efficiently computed without the interaction.
Such protocols play a central role in cryptography.  However,
practical protocols must be both secure and round-efficient.  Parallel
composition is a common technique for reducing the error probability
of an interactive protocol without increasing the number of rounds,
and therefore one is interested in parallel-composing {$\ZK$} protocols
while maintaining the {$\ZK$} property.  However, Goldreich and Krawczyk
\cite{GoldreichKrawczyk90} proved that only $\BPP$ languages have
three-round interactive proofs with negligible soundness error, that
are {\em black-box-simulation} {$\ZK$}.  This precludes parallel
composition of the well-known three-round {$\ZK$} protocols for Graph
Isomorphism while maintaining
black-box zero knowledge, unless the language is in {$\BPP$}.  Moreover,
\cite{GoldreichKrawczyk90} also precludes parallel composition of any
constant-round Arthur-Merlin ({$\AM$}) black-box-simulation {$\ZK$}
protocols except for languages in {$\BPP$}.   

Precise definitions of these terms, and of the other classes that we will informally introduce in this section, are given in \secref{s:preliminaries}.  Roughly, the concept of zero-knowledge is formalized by requiring an efficient {\em simulator} that produces a probability distribution indistinguishable from the distribution of the original verifier's conversations with the honest prover.  Black-box-simulation {$\ZK$} means that the simulator is only allowed to call the verifier as a black-box subroutine.  In an {$\AM$} protocol, the verifier's messages are fair coin tosses. 

\smallskip

In this work, we revisit the problem of parallel composition of black-box-{$\ZK$} protocols from the perspective of quantum computation, and find that the impossibility results of~\cite{GoldreichKrawczyk90} extend even to certain quantum cases.  Quantum computation has significant consequences for cryptography, especially since exponential speedups by quantum computers have been found for problems that are crucial in current cryptographic systems.  In the specific context of zero knowledge, quantum computers raise several interesting questions:
\begin{enumerate}
\item {\bf Quantum simulators:}
What happens if one weakens the zero-knowledge requirement to say that, if $x \in L$, anything efficiently computable after interacting with the prover, could also have been efficiently computed on a quantum computer without the interaction?  In other words, we allow the black-box simulator to be a quantum computer and ask if round-efficient {$\ZK$} protocols can exist for a larger class of languages than {$\BQP$} (refer to~\defref{t:blackboxdef}).  It is encouraging that black-box quantum simulators are known to be more powerful than black-box classical simulators in some settings.  For example, Watrous~\cite{Watrous06zk} has given a black-box quantum simulator for the standard three-round Graph Isomorphism protocol that succeeds with probability exactly one, whereas classical simulators for the same protocol succeed with probability only approaching one.  Perhaps quantum {\em exact} simulators, as in~\cite{Watrous06zk}, could be helpful in maintaining black-box {$\ZK$} under parallel composition.
\item {\bf Quantum messages:}
What happens for protocols with quantum messages?  We know that every quantum statistical zero-knowledge ({$\QSZK$}) language has a black-box quantum-simulation zero-knowledge, three-round quantum Arthur-Merlin ({$\QAM$}) protocol~\cite{Watrous02szk,Watrous06zk}.\footnote{The first and third messages of the {$\QAM$} protocol are quantum, and the second message, from the verifier, is a classical coin flip.  See \defref{t:am3def}.}  The soundness error of these protocols is exponentially close to $1/2$.  If the \cite{GoldreichKrawczyk90} result extends to the {$\QAM$} case, then this would give strong evidence against parallel repetition of {$\QAM$} protocols to reduce soundness error to be exponentially small, unless $\BQP = \QSZK$.
\end{enumerate}

\subsubsection*{Our Results:} We answer the first question above and
make partial progress on the second.  We prove that only $\BQP$
languages have three-round interactive protocols ($\IP$)
(see \thmref{t:ip3}), or constant-round $\AM$ protocols (see
\thmref{t:constantround}), that have negligible soundness error and
are black-box quantum simulation {$\ZK$}.\footnote{As every $\BQP$
language has a zero-round protocol with a quantum verifier, which is
trivially quantum-simulation black-box {$\ZK$}, this result
characterizes the class $\BQP$.}  Our results also hold if
the last message from the prover in these protocols is a quantum message.
In particular, only $\BQP$ languages have black-box quantum-simulation
{$\ZK$}, negligible-soundness, three-round $\QAM$ protocols with the
first two messages being classical.  
We show our results for {\em
computational} zero knowledge and therefore they apply as well for the
stricter notions of {\em statistical} and {\em perfect} zero
knowledge.  

\subsubsection*{Our Techniques:}

Let us now briefly discuss our techniques and the central idea of reduction to search.  For simplicity, assume a three-round {$\QAM$} protocol $\Pi$ for a language $L$ with all three protocol messages being classical but a quantum verifier (see \defref{t:am3def}).  Assume that $\Pi$ is black-box-simulation {$\QCZK$} with negligible soundness error. We prove $L \in \BQP$ by exhibiting an efficient quantum algorithm $\cZ$ that decides the language $L$.  Even though a similar algorithm works in the classical case studied by Goldreich and Krawczyk, our analysis of $\cZ$ is quite different from the analysis in~\cite{GoldreichKrawczyk90}.  For comparison, we therefore sketch the idea of the algorithm and of its analysis in this section.  The formal details appear in~\secref{s:threeroundamprotocols}.

Throughout the paper, we use capital letters to represent random variables, and lower-case letters to represent individual strings.  For a random variable $\a$, we let $\a$ also represent its distribution.

\begin{quote}
{\bf Idea of the algorithm $\cZ$:} Let $x$ be the input whose membership in $L$ needs to be decided. Since the protocol $\Pi$ is $\QCZK$, there exists a simulator $\cS$ with running time $t$ polynomial in $\abs{x}$.  Let $\h$ be a random variable uniformly distributed in $\bH(2t+1)$, where $\bH(2t+1)$ is a {\em strongly $(2t+1)$-universal family} of efficiently computable hash functions from $\{0,1\}^\na$ to $\{0,1\}^\nb$, where $\na, \nb$ are the lengths of the first and second messages, respectively, in $\Pi$ (see \defref{t:stronglyuniversalhash}).  For $h \in \bH(2t+1)$, let $\V_h$ represent a verifier who, if the first message is $\alpha$, replies with $h(\alpha)$.  Run $\cS$ on the random verifier $\V_\h$ and measure $\cS$'s output in the computational basis to obtain the (random) transcript $(\a,\b,\c)$; representing the prover Merlin's first message, the verifier Arthur's response and Merlin's second message, respectively.  Run Arthur's acceptance predicate on the modified transcript $(\a, \h(\a), \c)$, and declare $x \in L$ if and only if it accepts.
\end{quote}  

We claim that $\cZ$ accepts inputs $x \in L$, and rejects inputs $x
\notin L$, with good completeness and soundness parameters (see
\defref{t:am3def}).   

\noindent{\bf Sketch of proof:}
For $x \in L$, by using the zero-knowledge property of $L$ and
properties of the family of hash functions $\bH(2t+1)$, it can be
verified that the algorithm $\cZ$ accepts with good probability.  We
do not elaborate this case here. Instead we focus on the more
interesting case of $x \notin L$.  We show that if the algorithm
$\cZ$ accepts a string $x \notin L$ with probability $\epsilon$, then there
exists a cheating Merlin who fools the honest Arthur with probability
$\Omega(\epsilon/t^2)$.  This contradicts the protocol's soundness
being non-negligible for $\epsilon$ constant and $t$ polynomial.\footnote{In
the classical case, the cheating Merlin's success probability is
$\Omega(\epsilon/t)$, so a quantum black-box simulator can be no more
than quadratically more efficient.}

\def\aa {{\a'}}
\def\bb {{\b'}}
\def\cc {{\c'}}

The cheating Merlin $\M^*$ is designed as follows.  Since the algorithm $\cZ$
accepts $x \notin L$ with probability $\epsilon$, the modified
transcript $(\a,\h(\a),\c)$ satisfies Arthur's acceptance predicate
with probability $\epsilon$.  Therefore, a natural intention of
$\M^*$ could be to act so that the transcript of the actual
interaction is distributed ``close'' to $(\a,\h(\a),\c)$.  
$\M^*$ can start by sending the first message $\a'$ ($\a'
\in \{0,1\}^\na$), 
such that $\a'$ is distributed identical to $\a$.  Now Arthur, being
honest, replies with message $\b'$ uniformly distributed in
$\{0,1\}^\nb$  and  independent of $\a'$.  Now, we cannot show that the distribution of
the first two messages $(\a',\b')$ is either the same, or even close
in $\ell_1$ distance to the distribution of $(\a,\h(\a))$.  In
particular, $\h(\a)$ is not necessarily independent of $\a$.

However, using properties of the family $\bH(2t+1)$, 
we will argue below that $\h(\a)$ is ``well spread out,"
i.e., has sufficiently high {\em min-entropy}\footnote{For
a distribution $X$ taking values in $\cX$, min-entropy of $X$ is
defined to be $\min_{x \in \cX} - \log \Pr[X=x]$.} even conditioned on the value
of $\a$.  This means that $(\a', \b')$ can be ``closely coupled" to
$(\a, \h(\a))$.  For two distributions $P$ and $Q$, by saying that $P$
can be closely coupled to $Q$, we mean that the probabilities of $Q$,
scaled down by $t^2$, are point wise less than the corresponding
probabilities of $P$.  Note that then if a predicate accepts $Q$ with
probability $\epsilon$, it also accepts $P$ with probability
$\epsilon/t^2$.

Let us define random variable $\c'$ such that for all $\alpha \in
\{0,1\}^\na, \beta \in \{0,1\}^\nb, (\c' \vert(\a'=\alpha, \b' =
\beta)) = (\c \vert(\a=\alpha, \h(\alpha) = \beta))$. If the first
and second messages are $\alpha, \beta$ respectively, then    
$\M^*$ sends the third message distributed according to $\c' \vert (\a'= \alpha,
\b'= \beta)$. Due to this strategy of $\M^*$, the transcript of the actual
interaction $(\a', \b', \c')$, remains closely coupled to the
modified simulated transcript, $(\a, \h(\a), \c)$.  
Since we have assumed that the modified simulated transcript
$(\a, \h(\a), \c)$ satisfies Arthur's acceptance predicate with
probability $\epsilon$, from property of closely coupled distributions
that we mentioned above, Arthur is fooled to accept the actual
transcript $(\a', \b', \c')$ with probability at least
$\epsilon/t^2$. 

Since $\b'$ is uniform and independent of $\a'$, in order to show that
$(\a', \b')$  can be ``closely  coupled" to $(\a,  \h(\a))$, it can be
verified that it is enough to show that  $\h(\a)$ has high min-entropy
even conditioned on  the value of   $\a$.  Indeed, the  main technical
lemma of our paper,
\lemref{t:mainlemma}, shows that the simulator $\cS$, which can be
thought of as making at most $t$ queries to $\h$ and outputting $\a$ (in which
case $H,A$ become correlated random variables), cannot cause
$\h(\alpha)$ to have high min-entropy for most $\alpha$ distributed according to
$\a$.
By definition of min-entropy, this means for most $\alpha$, for any
$\beta$, the probability $\Pr[\h(\alpha) = \beta \vert \a = \alpha]$
is small.  In order to provide some intuition, let us assume $\cS'$ is
some classical algorithm making at most $t$ queries to a random
function $F$, chosen uniformly from the set of all functions from
$\{0,1\}^\na$ to $\{0,1\}^\nb$ and outputting $\a \in \{0,1\}^\na$.
We show the following weaker
statement; that is for all $\beta \in \{0,1\}^\nb$,
\begin{equation} \label{e:quantifierswitchedaround}
\Pr[F(\a) = \beta] \quad \leq \quad \frac{t+1}{2^\nb}.
\end{equation}
Let us fix a $\beta$.  The goal of $\cS'$ is now to maximize $\Pr[F(\a) =
\beta]$.  This can be viewed as a search problem.  It is easy to see that the
optimal procedure for $\cS'$ is:
\begin{quote}Make $t$ different queries to $F$.  If any response is
$\beta$ then output the corresponding queried location.  Otherwise,
output any new location.\end{quote}
Eq.~\eqnref{e:quantifierswitchedaround} is now immediate.  Note that,
since $\cS'$ makes at most $t$ queries to $\f$, this procedure would
also be optimal with the same probability of success even if $F$ were
only drawn uniformly from a strongly $(t+1)$-universal family of hash
functions.  In \lemref{t:mainlemma}, since $\cS$ is a quantum
algorithm and we need to show a stronger statement, the proof takes a
different track.  However, it also uses a reduction to the black-box
search problem.  
\qed 

Here, we would like to point out the main differences between our
analysis and the analysis in~\cite{GoldreichKrawczyk90}:
\begin{enumerate}
\item
The algorithm in
\cite{GoldreichKrawczyk90} constructs the responses of a random
function on the fly, as queries from $\cS$ to verifier arrive.
Quantumly, however since $\cS$ is a $\BQP$\ machine, queries can come in
superposition, and it is difficult to reply to them as a consistent,
uniform random function F, i.e., map $\sum_x \alpha_x \ket{x} \mapsto
\sum_x \alpha_x \ket{x} \ket{F(x)}$.  It is not even possible to
sample efficiently from the set of all functions from $\{0,1\}^\na$ to
$\{0,1\}^\nb$, since $\na, \nb$ are polynomial in $|x|$.  This is why
we must use a random hash function $\h$ drawn uniformly from
$\bH(2t+1)$, which is a much smaller family. However since $\h$ still
has $(2t+1)$-wise independence, it suffices for our purposes. 
\item
The more important difference is that~\cite{GoldreichKrawczyk90}'s arguments, showing that if their algorithm accepts an $x \notin L$ with good probability then there exists a good cheating prover, are essentially combinatorial. They can be phrased as inserting the honest Arthur into a random query round of the simulator.  Our arguments however cannot rely just on classical combinatorics, and a careful rephrasing (as sketched above) is needed to reduce the analysis to quantum search lower bounds. Since for the purpose of efficiency, we are forced to provide the input to the search algorithm, from a source of limited independence, a technical contribution of this work is also in showing that search is hard on average for such inputs as well.
\end{enumerate}

We would like to clarify one more aspect of the algorithm $\cZ$.  Why does $\cZ$ use $\V_H$, instead of running the simulator $\cS$ on the honest Arthur?  The reason is that the zero knowledge property of $L$ only restricts $\cS$'s behavior for $x \in L$.  However as we argued above, for $x \notin L$, we still want to be sure that $\cS$'s output has high min-entropy, even conditioned on its first message.  Using an efficiently computable hash function as a verifier in the algorithm $\cZ$, gives us some control on $\cS$'s output even when $x \notin L$; we can guarantee that the second message in $\cS$'s output is correct, and therefore not too concentrated.  Using a hash function works for the $x \in L$ case too, because the transcript of interaction with $\V_\h$ (averaged over randomness in $\h$) is distributed the same as the transcript with the honest Arthur.

Finally, the generalization to constant-round $\AM$ protocols goes through along similar lines.  These arguments also go through for three-round interactive protocols, by running the simulator on deterministic verifiers that use as their (private) random coins the hash of the prover's first message.

\subsection{Organization}

We make the necessary definitions including of our models in
\secref{s:preliminaries}.  In
\secref{s:threeroundamprotocols}, we give the proof for three-round 
$\QAM$ protocols.  We then generalize this proof in two directions.
First, we extend its validity to three-round quantum
interactive $\QIP$ (private-coins) protocols in~\secref{s:ip3}.  Next,
in \secref{s:constantroundamprotocols} we generalize it 
to constant-round $\QAM$ protocols, requiring slightly more involved
notation.  In \secref{s:openproblems} we conclude with some open problems.

\section{Preliminaries} \label{s:preliminaries}

We call a function $\delta$ {\em negligible}, $\delta \in \negl(n)$,
if for every positive polynomial $p$, $\delta(n) = O(1/p(n))$.  Let
$\poly(n)$ denote the set of functions that are each $O(p(n))$ for
some polynomial $p$.  We call an algorithm efficient if it can be run
on a classical or quantum Turing machine (depending on the context)
whose running time is at most polynomial in the input length.

We often use the following brief notation.  
Say $X_1$ and $X_2$ are random variables taking values in $\mathcal{X}$.  Let $x_1$, $x_2$ represent elements of $\mathcal{X}$.  
Then we write, for example, $\Pr[X_1 = X_2]$ to mean $\Pr_{(x_1, x_2) \leftarrow (X_1, X_2)} [x_1 = x_2]$. For better familiarity with the usual conventions and notations concerning random variables and other concepts of probability theory please refer to~\cite{Chung00}.

\subsection{Quantum Oracle}

\begin{definition} \label{t:quantumoracle}
A quantum oracle $U_f$ for a function $f: \{0,1\}^\na \rightarrow \{0,1\}^\nb$ is the unitary taking
\begin{equation} \label{e:quantumoracle}
\ket{x} \ket{a} \rightarrow \ket{x} \ket{a \oplus f(x)} \enspace, 
\end{equation}
for any $x \in \{0,1\}^\na$ and $a \in \{0,1\}^\nb$.  Here, $\oplus$
is the bitwise exclusive-or operation.  
\end{definition}
Note that $U_f$ is its own inverse, so oracle access to $U_f$ and
$U_f^{-1}$ is no more powerful than oracle access to just $U_f$. 

Below we provide brief definitions of classical and quantum
Interactive Proofs, Arthur-Merlin protocols, Zero-knowledge protocols
etc. For more detailed and precise definitions please refer
to~\cite{Goldreich99,GoldreichOren93,Watrous02szk,Watrous06zk}.
\subsection{Interactive proofs ($\IP$) and Arthur-Merlin protocols ($\AM$)}
A classical interactive proof ($\IP$) for a language $L$ is a classical communication
protocol between two parties, the prover $\P$ and the verifier $\V$.
Both parties receive the input $x$.  They exchange messages, and the
verifier finally outputs ``accept" or ``reject."  The verifier $\V$'s
running time is bounded by a polynomial in the length of $x$, but
there are no efficiency constraints on $\P$.  The protocol should
satisfy completeness and soundness requirements for some constants $\epsilon_c,
\epsilon_s > 0$ with $\epsilon_c + \epsilon_s < 2/3$:
\begin{enumerate}
\item
If $x \in L$, then the verifier $\V$ accepts with probability at least $1-\epsilon_c$.  
\item
If $x \notin L$, then no cheating prover $\P^*$ can make $\V$ accept
with probability more than $\epsilon_s$.   
\end{enumerate}

An $\AM$ protocol is a special kind of interactive proof in which the
verifier's messages are restricted to be uniformly random
coin flips, which are independent of each other and of prover's messages.

\subsection{Quantum Arthur-Merlin protocol ($\QAM$)} 

\label{s:am3def}

Similar to $\IP$ and $\AM$, we can also define quantum analogs, $\QIP$
and $\QAM$, where quantum messages are exchanged, and the verifier can
apply quantum operations.  For most parts in this paper, we are
concerned with special three-round quantum Arthur-Merlin ($\QAM$)
protocols in which only the third message, from the prover, is
quantum.  Therefore, we describe in detail only such protocols in
\defref{t:am3def} below. The details for the special three-round $\QIP$\ protocols
and special constant round $\QAM$\ protocols, with only the last message
being quantum, that we are also concerned
with in this paper, can be inferred easily from \defref{t:am3def} in an
analogous fashion. We begin with the following definition.

\begin{definition}[Quantum predicate] \label{t:qupredicatedef}
A quantum predicate is a two-outcome measurement given by an operator
$E$, $0 \leq E \leq I$.  When applied on a quantum state $\rho$, the
probabilities of the two outcomes, accept and reject, are $\tr E \rho$
and $\tr (I - E)\rho$, respectively.  The predicate is efficient if it
can be implemented in polynomial time by a quantum Turing machine.
\end{definition}

\begin{definition} [Special $\QAM$\ protocol]\label{t:am3def}
In a three-round quantum Arthur-Merlin ($\QAM$) protocol $\AMbr$ for
language $L$, with the first two messages being classical, verifier
Arthur ($\A$) and prover Merlin ($\M$) are each given the input $x \in
\{0,1\}^n$.  Then,  
\begin{enumerate}
\item Merlin sends Arthur an $\alpha \in \{0,1\}^{\na}$.  
\item Arthur replies with a uniformly random $\beta \in
\{0,1\}^{\nb}$, independent of the first message.   
\item Merlin sends $\rho$, a quantum state, and Arthur decides to
accept or reject based on an efficient quantum predicate (depending on
$x$) on the ``transcript" $\ketbra{\alpha}{\alpha} \tensor
\ketbra{\beta}{\beta} \tensor \rho$.    
\end{enumerate}
Here $n_1, n_2 \in \poly(n)$ and $\rho$ is a state on $\poly(n)$
qubits.  Note that there are no efficiency requirements on Merlin. For
convenience, we will let $(\alpha, \beta, \rho)$ denote the
transcript.  We will also write ``$\A$ accepts" to mean that Arthur's
predicate accepts.  Let $\AMbr(x)$ denote the distribution of protocol
transcripts $(\alpha, \beta, \rho)$ between Arthur $\A$ and Merlin
$\M$. We will also refer to $\AMbr(x)$ as the verifier's {\em view} in this protocol. 
The protocol satisfies, for some constants $\epsilon_c, \epsilon_s > 0$ with
$\epsilon_c + \epsilon_s < 2/3$: 
\begin{itemize}
\item {Completeness:} If $x \in L$, $\Pr(\textrm{$\A$ accepts
$\AMbr(x)$}) \geq 1-\epsilon_c$.   
\item {Soundness:} If $x \notin L$, then for any possibly cheating
Merlin $\M^*$, $\Pr(\textrm{$\A$ accepts $\langle \A,
\M^*\rangle(x)$}) \leq \epsilon_s$.   
\end{itemize}
\end{definition}
In the special three-round $\QIP$ protocols that we consider 
between prover $\P$ and verifier $\V$, the verifier's
view on input $x$ consists of its private coins together with
the transcript of the interaction.  We denote the random variable of
this view by $\langle\P,\V\rangle(x)$.

\subsection{Zero knowledge} 
Informally, as we have stated earlier, a zero-knowledge proof for a
language $L$ is an interactive proof for $L$ such that if $x \in L$,
then the verifier, no matter what it does, can ``learn nothing" more
than the validity of the assertion that $x \in
L$~\cite{GoldwasserMicaliRackoff89zero,GoldreichOren93}.  For a
cheating verifier $\V^*$, the notion of it not ``learning" more is
formalized, in the context of the protocols that we consider, 
using the definitions as follows.

\begin{definition}[Computationally indistinguishability]
\label{t:computationalindistinguishabilitydef} 
Two transcript distributions $A$  and $B$ on  $n$ classical or quantum
bits, are  said to be  \emph{computationally indistinguishable} if for
any efficient quantum predicate $M$ running in time polynomial in $n$,
\[
\left\vert \Pr[M \textrm{ accepts } A] - \Pr[M
\textrm{ accepts } B]  \right\vert \in \negl(n) \enspace . 
\]
\end{definition}

\begin{definition}[Quantum computational zero knowledge] \label{t:qczk}
An interactive protocol $\Pi$ (of the special kinds that
we consider) for language $L$, with prover $\P$ and
verifier $\V$, is computational zero knowledge if for every
efficient verifier $\V^*$ there exists an efficient quantum algorithm
$\cS^{\V^*}$, called the simulator, as follows.   
Let $\cS^{\V^*}(x)$ be $\cS$'s output on input $x$ representing
verifier's view in the protocol between $\P$ and $\V^*$.  Then for all $x \in L$, the
distributions of $\cS^{\V^*}(x)$ and of verifier $\V^*$'s actual view
$\langle\P,\V^*\rangle(x)$ while interacting with $\P$ are
computationally indistinguishable.   
\end{definition}

\begin{definition}[Black-box quantum computational zero knowledge] \label{t:blackboxdef}
An interactive protocol $\Pi$ (of the special kinds that we consider), is black-box quantum computational zero knowledge if there exists a single simulator $\cS$ that works for all efficient verifiers $\V^*$, and that uses the verifier $\V^*$ only as a black-box oracle.  That is, the access of $\cS$ to $\V^*$ is limited to querying $\V^*$ and receiving the response.
\end{definition}

The following remarks are in order: 
\begin{enumerate}
\item Perfect zero knowledge and statistical zero knowledge are two stronger
notions of zero knowledge that require the distributions of
$\cS^{\V^*}(x)$ and of $\langle\P,\V^*\rangle(x)$ to be the same or
{\em statistically indistinguishable}, respectively.  In the case of perfect
zero knowledge, the simulator is additionally allowed to output
``failure" instead of a transcript with probability $\leq 1/2$.
\item
Unlike the special quantum protocols that we consider, in which only the last message is quantum, 
for protocols with more quantum messages, the definition of quantum zero
knowledge needs changes.  For precise definitions,
please refer to~\cite{Watrous02szk, Watrous06zk}.  
\end{enumerate}

\section{Three-round $\QAM$ protocols with the first two messages classical} \label{s:threeroundamprotocols}

In this section we present our result for three-round $\QAM$\ protocols, \thmref{t:main}. 

\begin{theorem}
\label{t:main}
Let $L$ be a language with a three-round $\QAM$\ protocol $\Pi$ with the first two messages classical, as in \defref{t:am3def}, having completeness and soundness errors $\epsilon_c$ and $\epsilon_s$, respectively.  Assume that $\Pi$ is a black-box, quantum computational zero-knowledge protocol.  Let $\cS$ be the simulator with a running time bounded by $t$.  If $t \sqrt{\epsilon_s} = o(1 - \epsilon_c - \negl(n))$, then $L$ is in $\BQP$.
\end{theorem}

\noindent
In particular, if $\epsilon_s$ is negligible and $\epsilon_c$ a constant, then $L \in \BQP$.  

Although \defref{t:blackboxdef} requires a simulator that works for all efficient verifiers $\V^*$, the proof of \thmref{t:main} will only require that the simulator $\cS$ works for a limited set of verifiers, verifiers that essentially just apply a fixed function to the prover's message to determine their reply.  

\begin{definition}
For $h: \{0,1\}^\na \rightarrow \{0,1\}^\nb$, let $\V_h$ represent a dishonest verifier who replies deterministically $\beta = h(\alpha)$ on message $\alpha$, and uses the same acceptance predicate as used by Arthur $\A$.  
\end{definition}

In fact, in order for \thmref{t:main} to hold, the simulator $\cS$ only has to work for the set of cheating verifiers $\{ \V_h : h \in \bH \}$, where $\bH$ is a certain \emph{strongly $t$-universal family of hash functions}:

\begin{definition}[Strongly $t$-universal family of hash functions] \label{t:stronglyuniversalhash}
A set $\bH$ of functions from $\{0,1\}^\na$ to $\{0,1\}^\nb$ is a strongly $t$-universal family of hash functions if for $\h$ chosen uniformly from $\bH$, the random variables $\{\h(\alpha): \alpha \in \{0,1\}^\na\}$ are $t$-wise independent and each $\h(\alpha)$ is uniformly distributed in $\{0,1\}^\nb$.  

For all positive integers $\na, \nb, t$, there exists a strongly $t$-universal family $\bH(\na, \nb, t)$ of efficiently computable hash functions $\{0,1\}^\na \rightarrow \{0,1\}^\nb$~\cite{Joffe74, WegmanCarter81, ChorGoldreich89}.  
\end{definition}

With these definitions out of the way, we are ready to prove \thmref{t:main}.  

\begin{proof}[Proof of \thmref{t:main}]
The proof goes by presenting and analyzing the following efficient algorithm $\cZ$ for deciding membership in $L$:

\begin{center}
\fbox{
\begin{minipage}[l]{5.00in}
\noindent {\bf Algorithm $\cZ$:} Input $x \in \{0,1\}^n$, Output accept/reject.
\begin{enumerate}
\item Draw $\h$ uniformly from $\bH := \bH(\na, \nb, 2t+1)$.  
\item Run $\cS$ on $\V_\h$ with input $x$.  Consider the three output registers, corresponding to the prover's first message, the verifier's response, and the prover's second message, respectively.  In order to ensure that the first two messages in the simulated transcript are classical, measure the corresponding registers in the computational basis.  Let $\a$ and $\b$ be the respective random variables obtained after the measurement, and let $\c$ be the contents of the third register after the measurement.  Note that $\c$ is a random quantum state correlated with $\a$ and $\b$.  The output simulated transcript is then $(\a,\b,\c)$. 
\item 
Compute $\h(\a)$.  Run $\A$'s acceptance predicate on the modified simulated transcript $(\a,\h(\a),\c)$, and accept if and only if the predicate accepts.
\end{enumerate} 
\end{minipage} }
\end{center}

Algorithm $\cZ$ runs in polynomial time, since running $\cS$, choosing and evaluating a hash function in $\bH$, and running Arthur's acceptance predicate are all efficient.  We claim: 

\begin{claim} \label{t:algorithm}
For $x \in L$, $\Pr[\text{$\cZ$ accepts $x$}] \geq 1 - \epsilon_c - \negl(n)$.  For $x \notin L$, $\Pr[\text{$\cZ$ accepts $x$}] = O(t \sqrt{\epsilon_s})$.   
\end{claim}
\noindent \thmref{t:main} follows immediately from \claimref{t:algorithm}.    
\end{proof}

\def\abs #1{\lvert #1\rvert}

\begin{proof}[Proof of \claimref{t:algorithm}]

The first two steps of algorithm $\cZ$ define a joint distribution for $(\h, \a, \b, \c)$.  Here, $\a, \b$ are random variables taking values in binary strings, $\h$ is a random hash function, and $\c$ is a random density matrix.  Note that the algorithm does not use $\b$, the simulated second message.  $\cZ$'s acceptance probability is, from step 3,
\[
 \Pr[\text{$\cZ$ accepts $x$}] = \Pr[\text{$\A$ accepts $(\a, \h(\a), \c)$}] \enspace , 
\]
where the probability is over the joint distribution of $(\h, \a, \b,
\c)$, and also over any randomness in the acceptance predicate of
$\A$. 

\smallskip

\noindent {\bf Case $\mathbf{x \in L}$:} 
Let $x \in L$. Our aim is to relate  $\Pr[\cZ \textrm{ accepts } x]$ to $\Pr[\textrm{$\A$ accepts $\AMbr(x)$}]$, which is at least $1- \epsilon_c$ by the completeness criterion.  We compute
\begin{align} 
 \Pr[\text{$\cZ$ accepts $x$}] & = \Pr[\text{$\A$ accepts $(\a, \h(\a),
\c)$}]  \nonumber \\
&= \frac{1}{\abs{\bH}} \sum_{h \in \bH} \Pr[\textrm{$\A$
accepts $(\a,h(\a),\c)$} \vert \h = h] \nonumber \\
&\geq \frac{1}{\abs{\bH}} \sum_{h \in \bH}
\Pr[\textrm{$\A$ accepts $(\a,\b,\c)$ $\wedge$ $h(\a) = \b$} \vert \h
= h] \label{eq:xinL}
\end{align}
since $\h$ is uniform on $\bH$, and since adding the check $h(\a) = \b$ can only reduce the probability.   

Note that $(\a, \b, \c) \vert (\h = h)$ is the distribution of the simulator's output on verifier $\V_h$, after measuring the registers corresponding to the first two messages.  Let $\langle \V_h, \M\rangle(x)$ denote the distribution of protocol transcripts between verifier $\V_h$ and Merlin $\M$ on input $x$ (see \defref{t:am3def}).  By the computational zero-knowledge assumption, the acceptance probability of any efficient predicate on $(\a, \b, \c) \vert (\h = h)$ can differ from the acceptance probability of the same predicate on $\langle \V_h, \M\rangle(x)$ only by a negligible amount. In particular this holds for the following efficient quantum predicate $E$: on three-register input $\rho$, measure the first two registers, and accept iff $(\A$ accepts $\rho$ $\wedge$ $h(\text{first register}) = \text{second register})$.
Now, on $\langle \V_h, \M\rangle(x)$, the second message is by definition $h$ of the first message, so the event ($E$ accepts $\langle \V_h, \M\rangle(x)$) reduces to the event $(\A$ accepts $\langle \V_h, \M\rangle(x))$.  Therefore, continuing from Eq.~\eqref{eq:xinL} we have:
\begin{align*}
\Pr[\text{$\cZ$ accepts $x$}]
&\geq \frac{1}{\abs{\bH}} \sum_{h \in \bH}
\Pr[\textrm{$\A$ accepts $(\a,\b,\c)$ $\wedge$ $h(\a) = \b$} \vert \h
= h]  \\ 
& = \frac{1}{\abs{\bH}} \sum_{h \in \bH}
\Pr[E \textrm{ accepts } (\a,\b,\c) \vert \h = h]  \\
& \geq \frac{1}{\abs{\bH}} \sum_{h \in \bH} \Pr[E
\textrm{ accepts }\langle \V_h, \M \rangle(x)] - \negl(n) \\ 
& = \frac{1}{\abs{\bH}} \sum_{h \in \bH} \Pr[\textrm{$\A$
accepts $\langle \V_h, \M \rangle(x)$}] - \negl(n) \\  
&= \Pr[\textrm{$\A$ accepts $\langle \V_\h, \M \rangle(x)$}] - \negl(n) \enspace .
\end{align*}

Finally, since $\h$ is drawn from a strongly $(2t+1)$-universal hash family, for each $\alpha$, $\h(\alpha)$ is uniformly distributed.  Therefore, the transcript $\langle \V_\h, \M \rangle(x)$ is
distributed identically to $\AMbr(x)$; in either case, the second message is uniformly distributed and independent of the first message.  We conclude
\begin{align} \label{e:xinlaverage}
\Pr[\text{$\cZ$ accepts $x$}] & \geq  
\Pr[\textrm{$\A$ accepts $\langle \V_\h, \M \rangle(x)$}] - \negl(n)
\nonumber \\ 
& = \Pr[\textrm{$\A$ accepts $\AMbr(x)$}] - \negl(n)  \\
& \geq 1- \epsilon_c - \negl(n) \nonumber \enspace .
\end{align}

\smallskip

\noindent {\bf Case $\mathbf{x \notin L}$:} 
Let $x \notin L$. Let $q := \Pr[\cZ \textrm{ accepts } x] = \Pr[\textrm{$\A$ accepts $(\a,\h(\a),\c)$}]$.  Consider the following cheating Merlin $\M^*$.

\begin{center}\fbox{
\begin{minipage}[l]{5.00in}
\noindent {\bf Cheating Merlin $\M^*$}
Recall the joint distribution $(\h,\a,\b,\c)$ defined by $\cZ$. Note that $\h, \a$ need not be independent in this joint distribution.
\begin{enumerate}
\item On input $x$, send an $\alpha$ drawn according to $\a$.
\item On receiving $\A$'s message $\beta$, send back the quantum state $\c \vert (\a = \alpha, \h(\alpha) = \beta)$ to Arthur.  If $\Pr[\h(\alpha) = \beta \vert \a = \alpha] = 0$, then send state $\ketbra{0}{0}$.
\end{enumerate}
\end{minipage} }
\end{center}
Note that sampling from the conditional distribution $\c \vert (\a = \alpha, \h(\alpha) = \beta)$ may not be efficient.  However, $\M^*$ is not required to be efficient.  

The cheating probability of $\M^*$ is exactly 
\begin{multline} \label{e:cheatingprobability}
\Pr[\textrm{$\A$ accepts $\langle \A, \M^*
\rangle(x)$}] \\
= \sum_{(\alpha,\beta) \in \{0,1\}^{\na+\nb}} \Pr[\a = \alpha] \frac{1}{2^\nb} \Pr[\textrm{$\A$ accepts $(\alpha,\beta,(\c \vert \a = \alpha, \h(\alpha) = \beta))]$} \enspace . 
\end{multline}
The factor of $1/2^\nb$ is the probability with which Arthur replies with a given $\beta$.  By the soundness criterion, $\M^*$'s cheating probability is upper-bounded by $\epsilon_s$.  

Intuitively, $\M^*$ is only successful if the uniform distribution of $\beta$ has sufficient overlap with the distribution of $\h(\alpha)$ from the simulator's output, at least for most $\alpha$ drawn according to $\a$.  Then the two distributions can be coupled, relating Arthur's acceptance probability while interacting with $\M^*$ to $q$.  An extreme counterexample might be that conditioned on $\a = \alpha$; $\h(\alpha)$ were somehow fixed.  Then $\beta$ would almost never agree with $\h(\alpha)$, so $\M^*$ wouldn't know what to send for the last message and would have to abort.

Unlike the case $x \in L$, \defref{t:qczk} puts no guarantees on the simulator $\cS$ when $x \notin L$, so it is possible that $\cS$'s output $(\a, \b, \c)$ could be very different from $\AMbr(x)$.  Regardless, as we show in the following key lemma, one can argue using black-box query search lower bounds that $\h(\a)$ is on average not too concentrated even given $\a$. 

\begin{lemma}[Search reduction] \label{t:mainlemma}
Let $s_{\alpha} := \max_{\beta}\Pr[\h(\alpha) = \beta \vert \a=\alpha]$, where $(\h, \a, \b, \c)$ is the joint distribution defined in $\cZ$.  Then there is a universal constant $c$ such that the expectation 
\begin{equation} \label{e:mainlemma}
\Ex_{\alpha \leftarrow \a}[s_\alpha] \leq c t^2/2^\nb \enspace .
\end{equation}
\end{lemma}

\noindent
The proof is deferred to \secref{s:prooflowerboundreduction}.

By applying Markov's inequality to Eq.~\eqnref{e:mainlemma}, we obtain: 
\begin{corollary} \label{t:markov}
Fix $\delta \in (0,1]$.  There exists a set $\Good \subseteq \{0,1\}^\na$ such that:
\begin{enumerate}
\item $\Pr(\a \in \Good) \geq 1 - \delta$.
\item For all $\alpha \in \Good$, $s_\alpha \leq \frac{c t^2}{\delta 2^\nb}$.
\end{enumerate}
\end{corollary}

Now continuing from Eq.~\eqnref{e:cheatingprobability}, we have
\begin{align} \label{e:cheatingmerlin}
\epsilon_s & \geq \sum_{\alpha \in \Good, \beta} \Pr[\a = \alpha] \frac{1}{2^\nb} \Pr\!\big[\A \textrm{ accepts } (\alpha,\beta,(\c \big\vert \a = \alpha, \h(\alpha) = \beta))\big] \nonumber \\  
&= \sum_{\alpha \in \Good, \beta} \frac{ \Pr[\a = \alpha, \h(\alpha)=\beta]}{\Pr[\h(\alpha) = \beta \vert \a = \alpha]} \frac{1}{2^\nb} \Pr\!\big[\A \textrm{ accepts } (\alpha,\beta,\c) \big\vert \a = \alpha, \h(\alpha) = \beta \big] \nonumber \\ 
&\geq \frac{\delta}{c t^2} \sum_{\alpha \in \Good, \beta}{ \Pr[\a = \alpha, \h(\alpha)=\beta]} \Pr\!\big[\A \textrm{ accepts }(\alpha,\beta,\c) \big\vert \a = \alpha, \h(\alpha) = \beta\big] \enspace  \\
&= \frac{\delta}{c t^2} \Pr[\textrm{$\A$ accepts $(\a,\h(\a),\c)$, $\a
\in \Good$}] \nonumber \\ 
&\geq \frac{\delta}{c t^2} (\Pr[\textrm{$\A$
accepts $(\a,\h(\a),\c)$}] - \Pr[\a \notin \Good]) \nonumber \\ 
&\geq \frac{\delta}{c t^2} (q - \delta) \enspace . \nonumber
\end{align}
The second inequality above follows since $\Pr[\h(\alpha) = \beta \vert \a = \alpha] \leq s_{\alpha} \leq \frac{c t^2}{\delta 2^\nb}$, from the definition of $s_\alpha$, \lemref{t:mainlemma}, and since $\alpha \in \Good$. The final inequality uses the definition $q = \Pr[\text{$\cZ$
accepts $x$}] = \Pr[\textrm{$\A$ accepts $(\a,\h(\a),\c)$}]$ and 
\corref{t:markov}.  Set $\delta = q/2$ to complete the proof of \claimref{t:algorithm}, and
thus also of \thmref{t:main}. 
\end{proof}

\section{\texorpdfstring{Proof of \lemref{t:mainlemma}}{Proof of Lemma~\ref{t:mainlemma}}: Reduction to search} \label{s:prooflowerboundreduction} 

\lemref{t:mainlemma} is proved by reducing to search, then applying a search lower bound.  

We briefly sketch the idea of the proof first.  Let $s := \Ex_{\alpha \leftarrow \a}[s_\alpha]$, where $s_\alpha$ is as in the statement of the lemma.  For each $\alpha \in \{0,1\}^\na$, let 
$$
\beta_\alpha :=
\argmax_{\beta}\Pr[\h(\alpha) = \beta \vert
\a=\alpha] \enspace .
$$
(Recall the joint distribution of $(\h,\a,\b,\c)$ from algorithm $\cZ$.)  With this definition, note that $s = \Pr[\h(\a) = \beta_\a]$.  Let $(\a', \b', \c')$ be the simulator $\cS$'s output when run on $\V_F$, where $F$ is a uniformly random function from $\{0,1\}^\na$ to $\{0,1\}^\nb$.  Let $s' := \Pr[F({\a'})=\beta_{\a'}]$, where the probability is over both $F$ and the simulator.  First, we argue that $s' = s$ because the set of random variables $\{\h(\alpha): \alpha \in \{0,1\}^\na\}$ have sufficient independence.  Next, by reduction to black-box search and using known quantum search lower bounds, we argue that the probability of the event $(F({\a'})=\beta_{\a'})$ is $O(t^2/2^\nb)$ for any algorithm---in particular for $\cS$---that makes at most $t$ queries to oracle for $F$ and outputs $\a'$.  We now present the formal proof.

\begin{lemma} \label{t:twiseind}
Let $\h$ be uniformly distributed in $\bH(2t+1)$ and let $\f$ be uniformly distributed over the set of all functions $\{0,1\}^\na \rightarrow \{0,1\}^\nb$.  Let $\a = {\bf A}^\h \in \{0,1\}^\na$ be the classical output, after measurement, of a quantum algorithm ${\bf A}$ that starts in state $\ket 0$ and makes at most $t$ oracle queries to $\h$.  Let $\a' = {\bf A}^\f \in \{0,1\}^\na$ be the corresponding output when ${\bf A}$ is run on $\f$.  Then $(\a, \h(\a))$ and $(\a', \f(\a'))$ have the same distribution.  In particular, $\Pr[\h(\a) = \beta_\a] = \Pr[\f(\a') = \beta_{\a'}]$.
\end{lemma}

\begin{proof}[Proof of \lemref{t:twiseind}]
The proof follows by application of the polynomial
method~\cite{BealsBuhrmanCleveMoscaWolf98, NielsenChuang00}.  Given a
string $x = (x_\alpha)_{\alpha \in \{0,1\}^\na} \in \{ 0, 1\}^{\nb
2^\na}$, let $f_x: \{0,1\}^\na \rightarrow \{0,1\}^\nb$ 
be the function $f_x(\alpha) = x_\alpha$.  It is
well known that the state of the quantum query algorithm ${\bf A}$
starting at $\ket{0}$, after $t$ queries to the oracle for function
$f_x$ is $$
\sum_z p_z(x) \ket{z} \enspace ,
$$ 
where the coefficients $p_z(x)$ are polynomials in the binary
variables $x_{\alpha,i}$ with $\alpha \in \{0,1\}^\na$ and $i \in
[\nb] := \{1,\ldots, \nb\}$.  A block, for any fixed $\alpha$,
consists of the variables $x_{\alpha,i}$ with $i \in [\nb]$.  Also, it can
be verified through standard arguments,
 that each $p_z(x)$ has ``block degree" at most $t$,
meaning that each term involves variables $x_{\alpha,i}$ for at most
$t$ different $\alpha$: $$ p_z(x) = \sum_{\substack{d \leq t \\
\alpha_1, \ldots, \alpha_d \in \{0,1\}^\na \\ S_1, \ldots, S_d
\subseteq [\nb]}} p_{z,\alpha_1,\ldots,\alpha_d, S_1, \ldots, S_d}
\prod_{j=1}^d \prod_{i \in S_j} x_{\alpha_j,i} $$ 
Therefore, for a fixed $x \in \{ 0, 1\}^{\nb 2^\na}$, on making $t$
queries to the oracle for $f_x$, the probability of output of any
particular $\alpha$ is a polynomial of block degree at most
$2t$~\cite{BealsBuhrmanCleveMoscaWolf98}.  By making one additional
query to oracle for $f_x$, one can instead output $(\alpha,x_\alpha)$,
which increases the block degree by at most one.  That is, the
probability of output $(\alpha, x_\alpha)$ is a polynomial of block
degree at most $2t+1$. Averaging this polynomial over the oracle being
$\h$ gives the same probability as averaging over $\f$ by strong
$(2t+1)$-universality.  In either case, the variables $x_\alpha$ are
$(2t+1)$-wise independent and uniform.  Therefore, $(\a,
\h(\a))$ and $(\a', \f(\a'))$ have the same distribution.
\end{proof}

\begin{lemma} \label{t:mainlemmageneralsearch}
Let $\f$ be uniformly distributed over the set of all functions $\{0,1\}^\na \rightarrow \{0,1\}^\nb$.  Fix a sequence $(\beta_\alpha)_{\alpha \in \{0,1\}^\na}$ of elements of $\{0,1\}^\nb$.  Let $\a \in \{0,1\}^\na$ be the classical output, after measurement, of a quantum algorithm ${\bf A}$ that starts in state $\ket 0$ and makes at most $t$ oracle queries to $\f$.  
\begin{equation} \label{e:searchreduce}
\Pr[\f(\a) = \beta_\a] = O(t^2/2^\nb) \enspace .
\end{equation}
\end{lemma}

\begin{remark}
We state without proof that if ${\bf A}$ in \lemref{t:mainlemmageneralsearch} was a classical algorithm making at most $t$ oracle queries to $\f$, then we would have the stronger statement $\Pr[\f(\a) = \beta_\a] = O(t/2^\nb)$.
\end{remark}

\begin{proof}[Proof of \lemref{t:mainlemmageneralsearch}]
Let $S := \{x \in \{0,1\}^{2^\nb}: x \textrm{ has a $1$ in exactly one position} \}.$  Let $X$ be a random variable uniformly distributed in $S$.  Standard search lower bounds imply that with $t$ oracle queries to the bits of $X$, the probability of a quantum algorithm to find the location of the $1$ is $O(t^2/2^\nb)$~\cite{BealsBuhrmanCleveMoscaWolf98}.  (The same bound for a classical algorithm is $O(t/2^\nb)$.)

Now algorithm $\bf A$ can be used to construct an algorithm $\bf B$ for finding the $1$ in $X$ as follows:
\begin{quote}
Algorithm $\bf B$: Fix $G$ a function chosen uniformly from the set of all functions from $\{0,1\}^\na$ to $\{0,1\}^\nb$.  For each $\alpha \in \{0,1\}^\na$, fix $Z_\alpha$ a string chosen uniformly from $\{0,1\}^\nb \smallsetminus \{\beta_\alpha\}$.
Define, $$ F(\alpha) = \left\{\begin{split} \beta_\alpha \qquad
\text{if $X_{G(\alpha)} = 1$}\\  Z_\alpha \qquad \text{if
$X_{G(\alpha)} = 0$}\\\end{split}\right. \enspace.$$ Note that $F:
\{0,1\}^\na \rightarrow \{0,1\}^\nb$ is a uniformly random function, when averaged over the choices of $X, G$ and the $Z_\alpha$s.  Now run ${\bf
A}$.  When ${\bf A}$ makes a query to $\alpha \in \{0,1\}^\na$, return
$F(\alpha)$.\footnote{This response can be implemented in
superposition, using at most two oracle queries to $X$: choose
$\beta_\alpha$ or $Z_\alpha$ depending on $X_{G(\alpha)}$, then
uncompute $X_{G(\alpha)}$.  It is not necessarily efficient, except in
terms of oracle queries.}  When $\bf A$ stops, measure ${\bf A}$'s
output $A'$, and output $G(A')$.
\end{quote}
From the above construction, finding $\beta_\alpha$ in $F$
implies finding a $1$ in $X$.  Moreover, since $\bf A$ makes at most $t$ queries to $\f$, $\bf B$ makes at most $2t$ queries to $X$.  Therefore, 
\begin{equation*}
\Pr[F(\a') = \beta_{\a'}] = \Pr[X(G(\a')) = 1] = O(t^2/2^\nb) \enspace .
\end{equation*}
\end{proof}

\begin{proof}[Proof of \lemref{t:mainlemma}]
Recall the joint distribution of $(\h,\a,\b,\c)$ from the algorithm $\cZ$.  \lemref{t:mainlemma} now follows from above two lemmas by setting ${\bf A} := \cS$, $\beta_\alpha := \argmax_\beta \Pr[\h(\alpha) = \beta \vert \a = \alpha]$ and observing that $\Ex_{\alpha \leftarrow \a}[s_\alpha] = \Pr[\h(\a) = \beta_\a]$, where $s_\alpha$ is as in the statement of the \lemref{t:mainlemma}.
\end{proof}

\section{Three-round $\QIP$ protocols } \label{s:ip3}

The extension of \thmref{t:main} to a three-round 
interactive proof $\langle \V, \P \rangle$,  follows on similar lines as
the three-round $\QAM$\ case, with a few differences that we will
highlight.  Let us first introduce the notation for this section. 

{\bf{Notation:}}
In an interactive proof, the honest verifier $\V$ is given coins $R$ drawn uniformly at random from $\{0,1\}^{n_c}$ at the beginning of the protocol.  For a string $r$, let $\V^r : \{ 0, 1\}^\na \rightarrow \{0,1\}^\nb$ be the function determining verifier's $\V$'s response to the prover's first message, when the coins are fixed to $r$.  We will also write ``$\V^r$ accepts" to mean that $\V$'s predicate with coins $r$ accepts. For a function $h : \{0,1\}^\na \rightarrow \{0,1\}^{n_c}$, define the dishonest verifier $\V_h$ to behave exactly as the honest verifier $\V$ does with coins $h(\alpha)$, where $\alpha$ is the prover's first message.  In particular, $\V_h$ responds to message $\alpha$ with $\V^{h(\alpha)}(\alpha)$.
$\V_h$'s view of the interaction therefore consists of the two messages from the prover.   

The result for this section is: 

\begin{theorem} \label{t:ip3}
Let $L$ be a language with a three-round interactive protocol $\Pi$,
with possibly the last message from prover being quantum, having completeness and soundness errors $\epsilon_c$ and $\epsilon_s$, respectively.  Assume that $\Pi$ is a black-box, quantum computational zero-knowledge protocol.  Let $\cS$ be the simulator with a running time bounded by $t$.  If $t \sqrt{\epsilon_s} = o(1 - \epsilon_c - \negl(n))$, then $L$ is in $\BQP$.
\end{theorem}

\begin{proof}
The proof of \thmref{t:ip3} is similar to that of \thmref{t:main},
with some modifications to the algorithm and the cheating prover. The
new efficient algorithm $\cZ'$ for language $L$ is:

\begin{center}
\fbox{
\begin{minipage}[l]{5.00in}
\noindent {\bf Algorithm $\cZ'$:} Input $x \in \{0,1\}^n$, Output accept/reject.
\begin{enumerate}
\item Choose $\h$ uniformly at random from $\bH(2t+1)$.  Run $\cS$ (with input $x$) on
$\V_\h$ and measure its output corresponding to the first message from $\P$ in the computational basis to obtain the classical random variable $\a$. Let $C$ be the output of $\cS$ corresponding to the last message from $\P$. 
\item Accept if and only if $\V^{\h(\a)}$ accepts the transcript
$(\a,\V^{\h(\a)}(\a),\c)$. 
\end{enumerate} 
\end{minipage} }
\end{center}

\noindent
As before we have the following claim:
\begin{claim} \label{t:algorithmip}
For $x \in L$, $\Pr[\text{$\cZ'$ accepts $x$}] \geq 1 - \epsilon_c -
\negl(n)$.  For $x \notin L$, $\Pr[\text{$\cZ'$ accepts $x$}] = O(t
\sqrt{\epsilon_s})$.    
\end{claim}

\noindent
It is easy to verify that the algorithm $\cZ'$ runs in polynomial time.  \thmref{t:ip3} then follows immediately from \claimref{t:algorithmip}.
\end{proof}

\begin{proof}[Proof of \claimref{t:algorithmip}]
The case $x \in L$ goes along similar lines as in the proof of \thmref{t:main} and we skip the details for brevity.

Consider the case $x \notin L$.  Let $\b := \V^{\h(\a)}(\a)$.
Algorithm $\cZ'$ defines a joint distribution for $(\h, \a, \b, \c)$.
Let $\f$ be chosen uniformly from the set of all functions from
$\{0,1\}^\na$ to $\{0,1\}^{n_c}$.  Run the simulator $\cS$ on $\V_\f$
and let $\a', \c'$ be its outputs analogous to $\a, \c$.  Let $\b' :=
\V^{\f(\a')}(\a')$. Since $\cS$
makes at most $t$ queries, using arguments as in proof of~\lemref{t:twiseind} we have,
\begin{align} \label{eq:q}
q := \Pr[\text{$\cZ'$ accepts $x$}]
& = \Pr[\textrm{$\V^{\h(\a)}$ accepts $(\a,\b,\c)$}] \nonumber \\ 
& = \Pr[\textrm{$\V^{\f(\a')}$ accepts $(\a',\b',\c')$}] \enspace .
\end{align}

The main property that we need to observe in this case is:

\begin{lemma} \label{t:ip3lemma}
For all $\alpha \in \{0,1\}^\na$ and $\beta \in \{0,1\}^{\nb}$, the random variables $\f(\alpha) \vert (\a' = \alpha, \b' = \beta)$ and $\c' \vert (\a' = \alpha, \b' = \beta)$ are independent. In other words, for all $\alpha \in \{0,1\}^\na$, we have following Markov network:\footnote{The random variables $X, Y, Z$ taking values in $\cX, \cY, \cZ$ are said to form a \emph{Markov network} $X \rightarrow Y \rightarrow Z$ if for all $y \in \cY$ the random variables $X \vert (Y=y)$ and $Z \vert(Y=y)$ are independent.}
$$
(\f(\alpha) \vert \a' =  \alpha)   \rightarrow (\b' \vert \a' =
\alpha) \rightarrow (\c' \vert \a' = \alpha)
 \enspace . 
$$
\end{lemma}

\def\tV {{\mathbf{\tilde{V}}}}
\begin{proof}
Let $N:=2^{n_1}$. For every $\alpha \in \{0,1\}^\na$, define the random variable $Y(\alpha) := \V^{\f(\alpha)}(\alpha)$, so $\b' = Y(\a')$.  Note that the simulator $\cS$, while querying $\V_F$, has oracle access only to the
random function $Y : \{0,1\}^\na \rightarrow \{0,1\}^\nb$ and not
directly to the random function $\f : \{0,1\}^\na \rightarrow
\{0,1\}^\nc$.  Therefore the following is a Markov network:  
$$
(\f(1)\f(2)\ldots \f(N) ) \rightarrow (Y(1)Y(2) \ldots Y(N) ) \rightarrow 
(\a', \c') \enspace .
$$
The random variables $(\f(1)\f(2)\ldots \f(N))$ are all independent of each other.  Also, since for each $\alpha$, $Y(\alpha)$ is a function only of $\alpha$ and $\f(\alpha)$, the random variables $(Y(1)Y(2) \ldots Y(N))$ are also all independent of each other.  Therefore for every $\alpha \in \{0,1\}^\na$ we also have the following Markov network: 
$$
\f(\alpha) \rightarrow Y(\alpha) \rightarrow (\a', \c')
 \enspace ,
$$
which remains a Markov network if we condition each variable on $\a' = \alpha$, as claimed.  
\end{proof}

Exactly on the lines of \lemref{t:mainlemma}, search lower bounds imply:

\begin{lemma} \label{t:mainlemmaip}
Let $s_{\alpha} := \max_{r \in \{0,1\}^\nc} \Pr[\f(\alpha) = r \vert \a'=\alpha]$, where $(\f, \a', \b', \c')$ is the joint distribution defined as above.  Then there is a universal constant $c$ such that
the expectation 
\begin{equation} \label{e:mainlemmaip}
\Ex_{\alpha \leftarrow \a'}[s_\alpha] \leq c t^2/2^\nc \enspace .
\end{equation}
\end{lemma}

Applying Markov's inequality to Eq.~\eqnref{e:mainlemmaip} gives:

\begin{corollary} \label{t:markovip}
Fix $\delta \in (0,1]$.  There exists a set $\Good \subseteq \{0,1\}^\na$ such that:
\begin{enumerate}
\item $\Pr(\a' \in \Good) \geq 1 - \delta$.
\item For all $\alpha \in \Good$, $s_\alpha \leq \frac{c t^2}{\delta 2^\nc}$.
\end{enumerate}
\end{corollary}
Now define the cheating prover $\P^*$ as: 
\begin{center}\fbox{\begin{minipage}[l]{5.00in}
\noindent {\bf Cheating prover $\P^*$:}
Recall the joint distribution $(\f, \a', \b', \c')$ defined earlier.
\begin{enumerate}
\item On input $x$, send $\alpha$ drawn from $\a'$.
\item On receiving the honest verifier $\V$'s message $\beta$, 
send back message  to $\V$, distributed according to $\c' \vert(\a' = \alpha,
\b' = \beta)$.
\end{enumerate}
\end{minipage} }\end{center}

Now the cheating probability of $\P^*$ is $\Pr[\V \textrm{ accepts } \langle \V, \P^* \rangle(x)]$. Therefore, 
\begin{align*} \label{e:cheatingprobability1}
\epsilon_s & \geq \Pr[\V \textrm{ accepts } \langle
\V, \P^* \rangle(x)] \\ 
& = \sum_{(\alpha,r) \in \{0,1\}^{\na+\nc}} \Pr[\a' = \alpha] \frac{1}{2^\nc} \Pr[\V^r \textrm{ accepts } (\alpha,\V^r(\alpha),(\c' \vert \a' = \alpha, \b' = \V^r(\alpha)))] \enspace \\ 
& \geq \sum_{\alpha \in \Good, r} \Pr[\a' = \alpha] \frac{1}{2^\nc}
\Pr[\V^r \textrm{ accepts } (\alpha,\V^r(\alpha),(\c' \vert \a' = \alpha, \b' = \V^r(\alpha)))]  \\   
&= \sum_{\alpha \in \Good, r} \frac{ \Pr[\a' = \alpha, \f(\alpha)= r]}{\Pr[\f(\alpha) = r \vert \a' = \alpha]} \frac{1}{2^\nc} \Pr[\V^r \textrm{ accepts } (\alpha,\V^r(\alpha),(\c' \vert \a' = \alpha, \b' = \V^r(\alpha)))] \nonumber \\ 
&\geq \frac{\delta}{c t^2} \sum_{\alpha \in \Good, r}{ \Pr[\a' = \alpha, \f(\alpha)=r ]} \Pr[\V^r \textrm{ accepts } (\alpha,\V^r(\alpha),(\c' \vert \a' = \alpha, \b' = \V^r(\alpha)))]  \enspace  \nonumber \\
&= \frac{\delta}{c t^2} \sum_{\alpha \in \Good, r}{ \Pr[\a' = \alpha, \f(\alpha)=r ]} \Pr[\V^r \textrm{ accepts } (\alpha,\V^r(\alpha),(\c' \vert \a' = \alpha, \f(\alpha) = r ))]  \\
&= \frac{\delta}{c t^2} \sum_{\alpha \in \Good, r} \Pr[\a' = \alpha, \f(\alpha) = r, \V^r \textrm{ accepts } (\alpha, \V^r(\alpha), (\c' \vert \a' = \alpha, \f(\alpha) = r ))]  \\ 
&= \frac{\delta}{c t^2} \Pr[\V^{\f(\a')} \textrm{ accepts } (\a', \b' ,  \c' ) , \a' \in \Good ] \\
&=  \frac{\delta}{c t^2}  (\Pr[\V^{\f(\a')} \textrm{ accepts } (\a',\b',\c')] - \Pr[\a' \notin \Good]) \nonumber \\ 
&\geq \frac{\delta}{c t^2} (q - \delta) \enspace . \nonumber
\end{align*}
The third inequality above follows since $\Pr[\f(\alpha) = r \vert a' = \alpha] \leq s_{\alpha} \leq \frac{c t^2}{\delta 2^\nc}$ (from \corref{t:markovip}, the definition of $s_\alpha$, and since $\alpha \in  \Good$). The third equality above follows since from \lemref{t:ip3lemma}, 
$$ (\c' \vert \a'
= \alpha, \b' = \V^r(\alpha)) = (\c' \vert \a'
= \alpha, \f(\alpha) = r , \b' = \V^r(\alpha) ) = (\c' \vert \a'
= \alpha, \f(\alpha) = r ) \enspace .
$$ 
The final inequality uses Eq.~\eqnref{eq:q} and \corref{t:markovip}.
Set $\delta = q/2$ to complete the proof of \claimref{t:algorithmip},
and thus also of \thmref{t:ip3}. 
\end{proof}

\begin{remark}While we extend the three-round $\QAM$\ proof to constant-round $\QAM$\ protocols, as in \secref{s:constantroundamprotocols}, this proof for three-round $\QIP$ protocols cannot be similarly extended.  The proof would only work for constant-round $\QIP$ protocols if the honest verifier were guaranteed to use independent randomness to determine his response in each round.  The proof breaks down if it refers to the same randomness for different messages.  In that case, the black-box simulator's output transcript need \emph{not} only depend on the verifier's messages.  It may depend directly on the randomness behind that message, and so the analog to \lemref{t:ip3lemma} would be false.
\end{remark}

\section{Constant-round $\QAM$ protocols with only the last message quantum} \label{s:constantroundamprotocols}

In this section, we extend \thmref{t:main} for three-round $\QAM$\ protocols to $(2k+1)$-round $\QAM$\ protocols with all but the last message classical, for $k$ any constant.  

\begin{theorem} \label{t:constantround}
Let $k$ be a fixed positive integer.  Let $L$ be a language with a $(2k+1)$-round $\QAM$\ protocol $\Pi$, with all but the last message classical, having completeness and soundness errors $\epsilon_c$ and $\epsilon_s$, respectively.  Assume that $\Pi$ is a black-box, quantum
computational zero-knowledge protocol.  Let $\cS$ be the simulator with a running time bounded by $t$.  If $(t^{2k}\epsilon_s)^{1/(k+1)} = o(1 - \epsilon_c - \negl(n))$, then $L$ is in $\BQP$.
\end{theorem}

\begin{proof}

\def\vh {{\vec{h}}}
\def\bvh {{\vec{\mathbf{h}}}}
\def\bva {{\vec{\mathbf{a}}}}
\def\bvb {{\vec{\mathbf{b}}}}
\def\bvalpha {{\vec{\mathbf{\alpha}}}}
\def\bvbeta {{\vec{\mathbf{\beta}}}}
\def\balpha {{\mathbf{\beta}}}

\def\h {H}
\def\a {A}

Assume without loss of generality that the first message is from the prover $\M$.  We will use the following notation.

{\bf{Notation:}}
For an indexed variable $x_i$, let $x_1^j$ denote the $j$-tuple $(x_1, x_2, \ldots, x_j)$.  Let $n_i$ be the length of the $i$th message in the protocol.  For each $i \in [k]$, let $\bH_i$ be a strongly $(2t+1)$-universal family of efficiently computable hash functions $\{0,1\}^{N_i} \rightarrow \{0,1\}^{n_{2i}}$, with $N_i = n_1 + n_2 + \cdots + n_{2i-1}$ and a $t$ to be specified later (\defref{t:stronglyuniversalhash}).  We will use $\alpha_1, \ldots, \alpha_{2k}$ to denote classical messages of the first $2j$ rounds.  For hash functions, $h_1^k := (h_1, \ldots, h_k) \in \bH_1 \times \cdots \times \bH_k$, let $\A_{h_1^k}$ represent the deterministic dishonest Arthur who returns $h_{i}(\alpha_1^{2i-1})$ as the $(2i)$th message when the transcript of the first $2i-1$ messages is $\alpha_1^{2i-1}$.

Black-box access is modeled by giving the simulator $\cS$ access to $k$ oracles, evaluating the $k$ hash functions (on arbitrary inputs).  The oracle $U_{h_i}$ takes 
$$
\ket{x} \ket{a} \ket{b} \rightarrow \ket{c} \ket{x} \ket{a \oplus h_i(x)} \ket{b} 
$$
Equivalently, the simulator can be given a single oracle that takes as input also the round number to apply the appropriate hash function.

Let the random variables $\h_i$ be uniformly and independently distributed over $\bH_i$.  $\h_1^k := (\h_1,  \ldots, \h_k)$.  Let $(\a_1, \a_3, \ldots, \a_{2k+1})$ be 
the simulator $\cS$'s output for Arthur's view, corresponding to the prover's messages only, when run on the random verifier $\A_{\h_1^k}$.  ($\a_1, \a_3, \ldots, \a_{2k-1}$ are random classical messages, while $\a_{2k+1}$ is a random density matrix.)  Let $\a_{2i} = \h_i(\a_1^{2i-1})$, so $\a_1^{2k+1} := (\a_1, \a_2, \ldots, \a_{2k+1})$.  Thus running
$\cS$ on  $\A_{\h_1^k}$ overall defines a joint distribution over $(\h_1^k, \a_1^{2k+1})$.  As in the three-round case, our algorithm $\cZ$ for deciding $L$ is: 
\begin{center}
\fbox{
\begin{minipage}[l]{5.00in}
\noindent {\bf Algorithm $\cZ$:} On input $x$, using the simulator $\cS$, sample from the distribution $\A_1^{2k+1}$. Accept if and only if the sampled message satisfies Arthur's acceptance predicate.
\end{minipage} }
\end{center}

\noindent 
Our main claim will be: 
\begin{claim} \label{t:algorithm1}
For $x \in L$, $\Pr[\text{$\cZ$ accepts $x$}] \geq 1 - \epsilon_c - \negl(n)$.  For $x \notin L$, $\Pr[\text{$\cZ$ accepts $x$}] = O((t^{2k}\epsilon_s)^{1/(k+1)})$. 
\end{claim}
\noindent
Algorithm $\cZ$ runs in polynomial time, since $\cS$, choosing and
evaluating various hash functions, and Arthur's acceptance predicate
are all efficient.  Therefore, \thmref{t:constantround} follows immediately
from \claimref{t:algorithm1}.   
\end{proof}

\begin{proof}[Proof of \claimref{t:algorithm1}]
Let
\begin{equation} \label{e:generalqdefinition}
q := \Pr[\text{$\cZ$ accepts $x$}] = \Pr[\text{$\A$ accepts $\a_1^{2k+1}$}] \enspace .
\end{equation}
For $x \in L$, $\cZ$ accepts with good probability by the computational zero knowledge assumption and by averaging over the hash functions, as in \thmref{t:main}.  
We skip the details for brevity and focus on the $x \notin L$ case.  Define a cheating Merlin $\M^*$ as follows: 
\begin{center}
\fbox{
\begin{minipage}[l]{5.00in}
\noindent {\bf Cheating Merlin $\M^*$:} If the transcript so far is
$\alpha_1^{2i}$, send the next message according to the
distribution of $\a_{2i+1} \vert (\a_1^{2i} = \alpha_1^{2i})$.  
\end{minipage} }
\end{center}
The cheating probability of $\M^*$ is $\Pr[\A \textrm{ accepts } \langle \A, \M^* \rangle(x)]$.  Therefore,
\begin{multline} \label{e:generalcheatingprobability} 
\epsilon_s  \quad \geq \quad \Pr[\A \textrm{ accepts }\langle \A, \M^* \rangle(x)] \\
 = \sum_{\alpha_1^{2k+1} \in \{0,1\}^{N_{k+1}}} 
\left(
\begin{split} 
\Pr[\a_1 = \alpha_1] \frac{1}{2^{n_2}} \Pr[\a_3 = \alpha_3 \vert
\a_1^2 = \alpha_1^2] 
\cdots \frac{1}{2^{n_{2k}}} \Pr[\a_{2k+1} = \alpha_{2k+1} \vert
\a_1^{2k} = \alpha_1^{2k}] \\ \times \Pr[\A \textrm{ accepts } \alpha_1^{2k+1}]  
\end{split} 
\right) \enspace .
\end{multline}

Let 
$$
\alpha_{2i}(\alpha_1^{2i-1}) := \argmax_{\alpha_{2i}} \Pr[\a_{2i} = \alpha_{2i} \vert \a_1^{2i-1} = \alpha_1^{2i-1}] \enspace .
$$  
Then using arguments involving search lower bounds as in~\lemref{t:mainlemmageneralsearch}, we can similarly conclude:
\begin{equation*}
\forall i \in [k]: \Pr[\a_{2i} = \alpha_{2i}(\a_1^{2i-1})] \leq
\frac{c t^2}{2^{n_{2i}}}
\end{equation*}
for some constant $c$.  Let $\delta \in (0,1]$.  By Markov's inequality, for all $i \in [k]$, there exists $\Good_i \subseteq \{0,1\}^{N_i}$, such that:
\begin{enumerate}
\item $\Pr[\a_1^{2i-1} \in \Good_i] \geq 1 - \delta$.
\item For all $\alpha_1^{2i-1} \in \Good_i$, $\Pr[\a_{2i} = \alpha_{2i}(\a_1^{2i-1}) \vert \a_1^{2i-1} = \alpha_1^{2i-1}] \leq \frac{c t^2}{\delta 2^{n_{2i}}}$.
\end{enumerate}
Let $\Good = \bigcap_{i=1}^{k} \Good_i \times
\{0,1\}^{N_{k+1}-N_i}$.  Then 
\begin{equation} \label{eq:markov}
\Pr[\a_1^{2k+1} \in \Good] \geq 1-k \delta \enspace .
\end{equation}
Now from Eq.~\eqnref{e:generalcheatingprobability}, we have:
\begin{multline*}
\epsilon_s
\geq \sum_{\alpha_1^{2k+1} \in \{0,1\}^{N_{k+1}}} 
\left(
\begin{split} 
\Pr[\a_1 = \alpha_1] \frac{1}{2^{n_2}} \Pr[\a_3 = \alpha_3 \vert
\a_1^2 = \alpha_1^2] 
\cdots \frac{1}{2^{n_{2k}}} \Pr[\a_{2k+1} = \alpha_{2k+1} \vert
\a_1^{2k} = \alpha_1^{2k}] \\ \times \Pr[\A \textrm{ accepts } \alpha_1^{2k+1}]  
\end{split} 
\right) \\
\begin{split}
&\geq \sum_{\alpha_1^{2k+1} \in \Good} 
\left( \begin{split}
(\tfrac{\delta}{c t^2} \Pr[\a_2 = \alpha_2 \vert \a_1 = \alpha_1]) \cdots (\tfrac{\delta}{c t^2} \Pr[\a_{2k} = \alpha_{2k} \vert \a_1^{2k-1} = \alpha_1^{2k-1}]) \\
\times \Pr[\a_1 = \alpha_1] \Pr[\a_3 = \alpha_3 \vert \a_1^2 = \alpha_1^2] 
\cdots \Pr[\a_{2k+1} = \alpha_{2k+1} \vert \a_1^{2k} = \alpha_1^{2k}] \\
\times \Pr[\textrm{$\A$ accepts $\alpha_1^{2k+1}$} ] \
\end{split} \right) \\
&= \left(\frac{\delta}{c t^2}\right)^k \Pr[\textrm{$\A$ accepts $\a_1^{2k+1}$, $\a_1^{2k+1} \in \Good$}] \\
&\geq \left(\frac{\delta}{c t^2}\right)^k (\Pr[\textrm{$\A$ accepts $\a_1^{2k+1}$}] - \Pr[\a_1^{2k+1} \notin \Good]) \\
&\geq \left(\frac{\delta}{c t^2}\right)^k (q - k \delta)
\enspace .
\end{split}\end{multline*}
The first inequality is by restricting the sum to good transcripts, and inserting terms $\tfrac{\delta 2^{n_{2i}}}{c t^2}\Pr[\a_{2i} = \alpha_{2i}(\a_1^{2i-1}) \vert \a_1^{2i-1} = \alpha_1^{2i-1}] \leq 1$.
(Compare to Eq.~\eqnref{e:cheatingmerlin}.)  
The last inequality follows from Eq.~\eqref{e:generalqdefinition} and 
Eq.~\eqref{eq:markov}.
Setting $\delta = q/(2k)$,  completes the proof. 
\end{proof}

\begin{remark}
This proof would not have gone through had we defined $\a_{2i} = \h_i(\a_{2i-1})$; it is necessary to hash the entire preceding transcript $\a_{2i} = \h_i(\a_1^{2i-1})$ (as in~\cite{GoldreichKrawczyk90}), in order to put an upper bound on $\Pr[\a_{2i} = \alpha_{2i} \vert \a_1^{2i-1} = a_1^{2i-1}] \leq \Pr[\a_{2i} = \alpha_{2i}(\a_1^{2i-1}) \vert \a_1^{2i-1} =
a_1^{2i-1}]$.
\end{remark}

\section{Open problems} \label{s:openproblems}

Many open problems remain related to this work.  We would like to be
able to analyze protocols with more ``quantum-ness."  For example,
what can one say about three-round interactive proofs with classical
messages but a quantum verifier?  Here the honest verifier may not
even have any private coins, but instead may use quantum mechanics to
randomize.  Since the verifier's response will no longer be a
deterministic function $\V^r(\alpha)$ of its coins $r$ and the first
message $\alpha$, our approach of setting the coins equal to a
function $h(\alpha)$ will not make sense.

Also, we would like to understand $\QAM$\ protocols in which all the
prover's messages are quantum.  The problem currently is that hashing the first
message (say, in the computational basis) collapses its state.
Therefore it is no longer true that the honest-verifier transcript is
the same as the average of the hash-function verifiers transcripts, so
the key equality in the $x \in L$ case, Eq.~\eqnref{e:xinlaverage},
will no longer hold.

\vspace{0.2in}
\noindent{\bf Acknowledgements} \newline
We appreciate many helpful discussions with Sean Hallgren, Umesh
Vazirani and Hoeteck Wee.  We thank the anonymous referees for helping us improve the presentation of the paper. 
We acknowledge support from NSF ITR Grant
CCR-0121555, and ARO Grant DAAD 19-03-1-0082.  The large part of this  work was conducted
while the authors were at the University of California, Berkeley. Part
of the work was conducted when the first author was at the School of Computer
Science and Institute for Quantum Computing, University of
Waterloo, where it was supported by an ARO grant.

\end{document}